\documentclass{emulateapj}

\usepackage{epsfig}
\usepackage{amsmath}
\usepackage{natbib}

\begin{document}

\title{The Submillimetre Universe\footnotemark[$\dagger$]}
\footnotetext[$\dagger$]{A White Paper submitted to the Canadian Long Range
Plan 2010}
\shorttitle{The Submm Universe}
\author{Douglas Scott, 
Pauline Barmby, 
Pierre Bastien,
Jan Cami, 
Edward Chapin, 
James Di Francesco, 
Michel Fich, 
Mark Halpern, 
Martin Houde, 
Gilles Joncas, 
Douglas Johnstone, 
Peter Martin, 
Gaelen Marsden, 
Brenda Matthews, 
David Naylor, 
C. Barth Netterfield, 
Els Peeters, 
Rene Plume, 
Alexandra Pope, 
Gerald Schieven, 
Tracy Webb, 
Christine Wilson} 
\shortauthors{Scott et al.}

\journalinfo{}
\submitted{16th February, 2010}

\begin{abstract}
Submillimetre continuum radiation allows us to probe cold objects,
particularly the earliest, dusty phases of star formation, high-redshift
galaxies and circumstellar disks.  The submillimetre window gives a unique
view of the physical and dynamical
conditions in the neutral and molecular interstellar medium.
In the next decade a combination of wide-field surveys with single-dish
telescopes and targeted follow-up with ALMA and other facilities
should enable rapid progress in answering questions about
the origins of planetary systems, stars and galaxies.
\end{abstract}
\keywords{Submm -- Universe}

\section{The Submm Window}

The sub-millimetre (hereafter `submm') waveband is usually defined to
stretch from $200\,\mu$m to $1\,$mm.  There are strong physical reasons
why this range of wavelengths is of particular astronomical interest,
reasons which are partly shared with the neighbouring far-IR and mm
bands.  In accord with the `Origins' theme of the 2000 Canadian Long Range
Plan, the earliest stages of planet, star and galaxy formation are most
directly amenable to study at submm wavelengths.

The youngest systems in the Universe tend to be shrouded in dust, which
absorbs shorter wavelength radiation and re-emits at hundreds of microns
(Fig.~1).
So to study cold systems, including
the early stages of star formation, debris disks in nearby
stellar systems and high redshift star-forming galaxies, one naturally
turns to submm wavelengths, where the opacity is low and thermal emission
is being probed directly.

\vspace{0.5truecm}
\begin{figure}
  \begin{center}
	\null\hspace{-0.4truecm}
    \leavevmode
      \epsfxsize=9cm\epsfbox{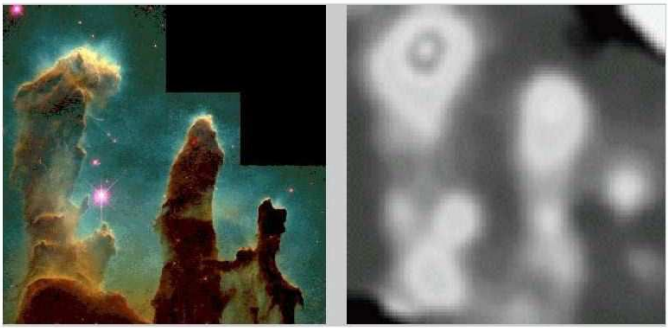}
       \caption{{\it Left\/}: Iconic optical image of the Eagle
       Nebula taken by the {\it Hubble Space Telescope}, showing the `elephant
       trunk' columns protruding from the molecular cloud, illuminated by
       nearby young stars, but with the youngest objects buried inside.
       {\it Right\/}: SCUBA image at $450\,\mu$m showing thermal
       dust emission, unveiling the cold cores where the earliest stages of
       star formation can be studied [1].}
     \label{fig:Eagle}
  \end{center}
\end{figure}

In addition to continuum radiation, spectral features in the submm are
diagnostics of the detailed physical conditions within dust and gas clouds.
In fact most known interstellar medium (ISM) molecules have their lowest
transitions at far-IR to mm wavelengths.  The
rich spectrum in the sub-mm, including dominant cooling lines of the atomic
and molecular components of the ISM,
traces a range of excitation properties.
Sub-mm spectroscopy determines density, temperature, etc.\ in ionized and
neutral gas, as well as dust.  This is the only part of the electromagnetic
spectrum which provides a complete picture of all phases of ISM.
And in extragalactic systems one can also learn about the hardness of
the interstellar radiation field, distinguishing black hole sources in
Active Galactic Nuclei (AGN) from hot young stars.

The relatively long wavelengths mean that single dish submm telescopes
have beamsizes typically ${\sim}\,10^{\prime\prime}$, which is a major
limitation for probing down to the scales most relevant for understanding
star formation, and also makes identification of extragalactic sources
challenging.  However, submm interferometers provide the
kind of resolution which is routine at optical wavelengths, and indeed are
able to reach below $0.1^{\prime\prime}$.  However, one must appreciate that
the field of view of an interferometer is determined by the size of each
antenna -- hence it is essentially the same as the {\it beamsize\/} of a
single dish telescope.  This makes it natural to use large-format array
detectors on single dish telescopes for surveys, combined with follow-up
observations with interferometers.

\begin{figure}
  \begin{center}
    \leavevmode
      \epsfxsize=9cm\epsfbox{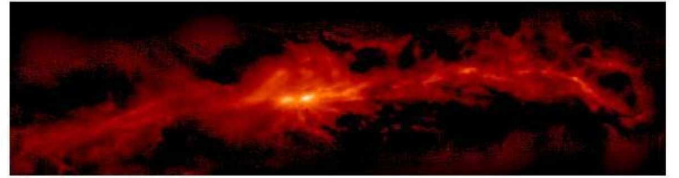}
       \caption{Part of the `integral-shaped filament' in Orion, imaged
       using the SCUBA camera on the JCMT [6].}
     \label{fig:integral}
  \end{center}
\end{figure}

\begin{figure*}
  \begin{center}
    \leavevmode
      \epsfxsize=19cm\epsfbox{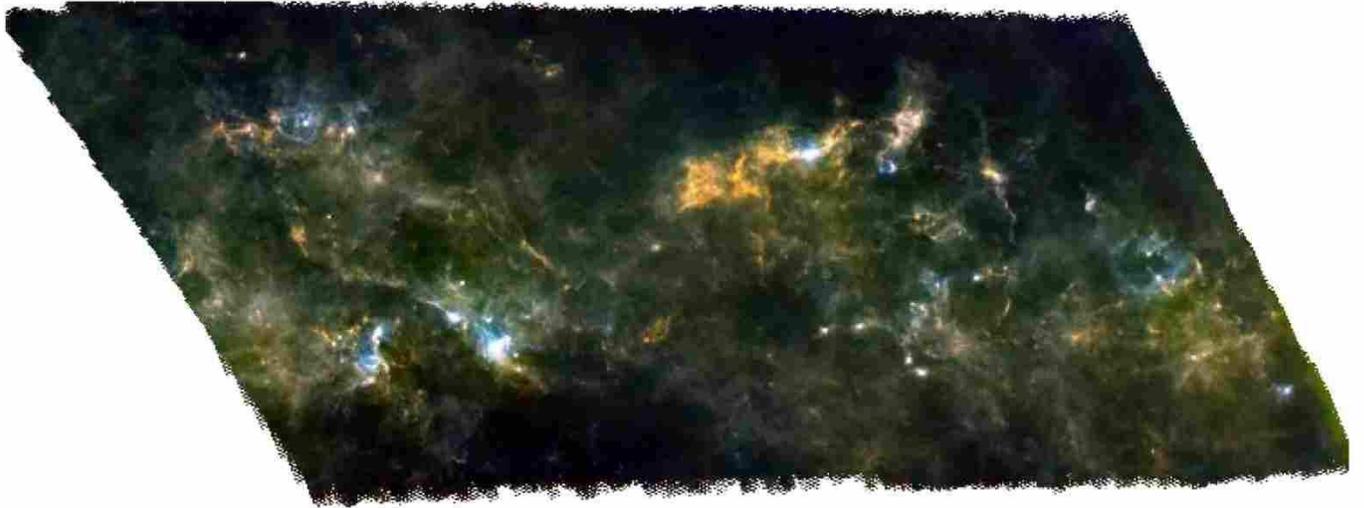}
       \caption{Submillimetre 3-band image of the Vela star-forming
       region [9].  The image covers about
       $50\,{\rm deg}^2$, and shows a wide range of morphologies associated
       with recent and on-going star formation.  The combination of filters
       spanning the thermal peak for cold sources means that one can literally
       {\it see\/} temperature in this image -- red regions are genuinely
       colder.}
     \label{fig:Vela}
  \end{center}
\end{figure*}

Polarization gives another dimension to submm studies, tracing
magnetic fields through alignment of dust grains.  Through continuum
measurements, spectroscopy and polarimetry
a wide range of science topics can be explored at submm wavelengths,
including studies of planetary atmospheres, comets, Kuiper Belt objects,
evolved stars and planetary nebulae.  But the highest impact has
come from results from extragalactic astronomy, star formation and
circumstellar disks -- all areas in which Canadians have made major
contributions.

\section{The Canadian Context}

In 1987 Canada joined the U.K. and the Netherlands in a partnership to run
the James Clerk Maxwell Telescope (JCMT [2]).
This effectively built on Canadian
expertise in atomic and molecular spectroscopy (personified in Gerhard
Herzberg) and radio astronomy (with telescopes at Penticton and Algonquin).
Now Canadian astronomers had access to the best submm facility in the world,
and they were quick to exploit it.  With early instrumentation consisting of
single element bolometers and heterodyne spectrometers, it was natural that
relatively bright Galactic objects were the main focus of study.  Canadian
users of the JCMT therefore made significant contributions to our
understanding of the ISM and star formation, as well as Solar System objects.

However, the arrival of the Submillimetre Common-User
Bolometer Array (SCUBA) in 1996 made it routine to map fainter objects, and
the submm window opened up to the extragalactic community as well.
In the years around 2000 the JCMT (and SCUBA in particular) achieved citation
rates which were second only to the {\it Hubble Space Telescope} among all
astronomical facilities (see [3]).  Specific Canadian contributions include 
the Canada-UK Deep Submillimeter Survey (e.g.~[4]) and imaging of
the Hubble Deep Field (e.g.~[5]) and Integral-shaped Filament in
Orion (Fig.~2 [6]), to focus on some of the higher impact results.

As well as observational and theoretical work, the submm has also been
a fruitful waveband for developing instrumental expertise within Canada.
Several hardware projects have only been possible through the existence of
the instrumentation group at the Herzberg Institute of Astrophysics.  But there
has also been experimental development at many Canadian universities, including
British Columbia, Lethbridge, McGill, Montreal, Toronto, Waterloo and Western
Ontario.

A vigorous Canadian programme in studying the Cosmic Microwave Background
(through rockets and balloons) also has a strong synergy with the science
goals of submm instruments (see [7]).
This explicitly resulted in the Balloon-borne
Large-Aperture Submillimetre Telescope (BLAST), a US-led project with strong
involvement from UBC and UofT.  A precursor to the SPIRE instrument on the
{\it Herschel\/} satellite, BLAST has had 2 successful flights, resulting in
more than 20 refereed papers (e.g.~[8], see Fig.~3).

Through a successful CFI grant, Canada became a partner in the development of
SCUBA-2, an ambitious submm array camera using ${\sim}\,10^4$ transition edge
sensor bolometers, described in more detail in a separate paper [2].
A series of community discussions and proposals resulted in the following
7 surveys
being allocated a large slice of JCMT telescope time: SCUBA-2 `All-Sky' Survey;
Cosmology Legacy Survey; Nearby Galaxies Survey; JCMT Galactic Plane Survey;
Gould Belt Survey; Spectral Legacy Survey; and Debris Disk Survey.
Although SCUBA-2 is late, the importance of the surveys remains high, and
the capabilities of the instrument are still unique\footnote{It
is worth remembering that neither SCUBA-2 nor BLAST were
mentioned in the 2000 LRP -- the importance of the ability to respond quickly
to special opportunities may not be unique to the submm band, but it has
certainly been critical in recent years.}.

The most substantial increase in the capabilities available for Canada at
submm wavelengths is the involvement in the Atacama Large
Millimetre/submillimetre Array (also described in a separate paper [10]).
Beyond the useful life of SCUBA-2, access to wide-field surveys could come
from participation in a new project like CCAT [11].

In space (see [12, 13])
Canadians are involved in both Guaranteed and Open Time projects
on {\it Herschel}, e.g.~HerMES and Hi-GAL, and Canadians are also working on
Galactic and extragalactic `foreground' science projects using {\it Planck\/}
data.  In addition Canada is playing a
role in plans for the SAFARI instrument on SPICA, which is expected to launch
in the middle of this decade.


\section{Galactic Science Goals}

\subsection{Star Formation}

Massive stars form from dense cores within molecular clouds, and
are important for chemical enrichment, cycling the
ISM into stars (see [14]).
A full grasp of galaxy evolution requires understanding
star formation in detail, but there are still many unsolved issues in the
star-formation problem.  The youngest stars form in dusty cores -- finding
them demands wide submm surveys, with high resolution continuum and
spectroscopic follow-up to probe accretion, outflows and disks.

There seems to be a similarity between the mass function of clumps and
the mass function of stars -- what precisely is that connection, and can we
understand the origin of the Initial Mass Function (IMF)?  This requires
measurements that
probe down to substellar masses and wide surveys to test the environmental
dependence of the IMF.  Detailed kinematic studies can only be done with
heterodyne mapping, also typically covering large areas (e.g.\ Fig.~4).

Details of the formation, structure and evolution of stars and stellar systems
are still poorly understood, largely because of the physical complexity,
involving accretion, atomic and molecular cooling, astrochemistry,
dynamics and magnetic fields.  Theoretical modelling struggles to keep pace
with the quality of the data, and observations must reach the scale of the
protostars themselves -- ultimately requiring space-based interferometers.
To observationally determine how star formation results in a diversity of
systems, including planets, we need a combination of
spatial resolution, spectral resolution, wavelength coverage, mapping
coverage and sensitivity.

\begin{figure}
  \begin{center}
    \leavevmode
      \epsfxsize=8cm\epsfbox{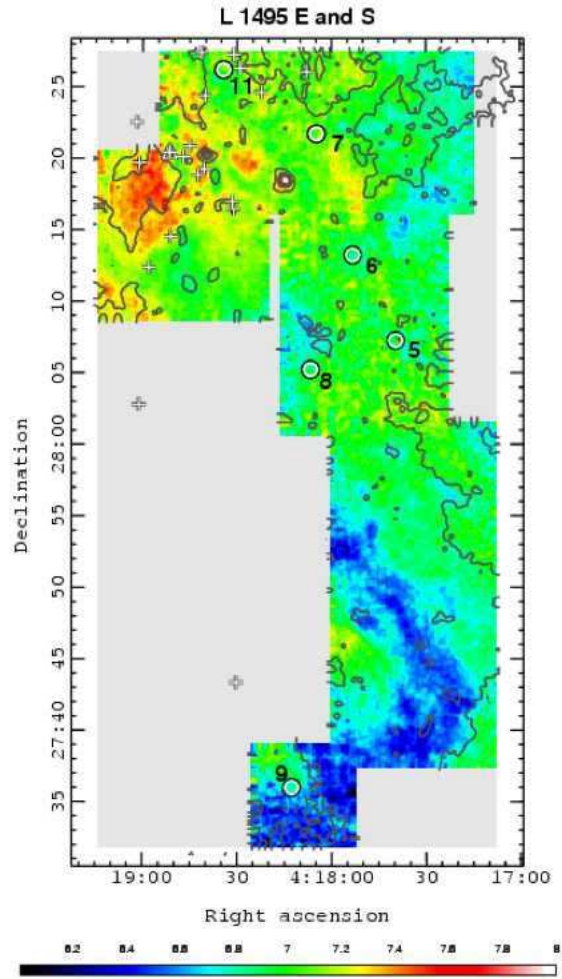}
       \caption{Recent example of a dynamical study of a low mass
       star-forming region: radial velocity map in Taurus,
       using CO$\,3{-}2$ data-cubes from the HARP
       instrument on JCMT [15].  The colour bar shows velocities
       in the range 6 to $8\,{\rm km}\,{\rm s}^{-1}$, while the contours show
       integrated CO intensity.  Circles are H$^{13}$CO$^+$ cores, crosses are
       Young Stellar Objects and triangles are Herbig-Haro objects.}
     \label{fig:Taurus}
  \end{center}
\end{figure}

\subsection{The Interstellar Medium}

For a complete picture of the Milky Way, one must
trace the cycle of molecules from the diffuse ISM to planetary systems,
again requiring wide-field mapping combined with
targeted spectroscopy.  Here digital spectroscopic
advances seem likely to revolutionize the field, leading perhaps to the
ultimate goal of connecting astrochemistry to astrobiology.

Astrochemistry is fundamentally about complexity -- getting the big picture
requires observations, analysis and modelling of many molecular and atomic
lines and their correlations.  This is also tied with the formation,
distribution, composition and evolution of dust grains.  We would like to
follow the full lifecycle of dust, from asymptotic giant branch stars,
supernovae and young stellar objects into the ISM and through to
the `death' of dust in star-forming regions.  The resolution of HIFI on
{\it Herschel\/} allows for detailed compositional and kinematic studies.
SPICA will have similar resolution, but much improved sensitivity.

\subsection{Polarization}

The role of magnetic fields in the diffuse ISM is key for tracking the assembly
of clouds and star-forming cores (Fig.~5)
and feedback into the ISM (Fig.~6).   Sub-mm polarimeters are complementary to
other measurement techniques (such as optical polarimetry and H{\sc i}
Zeeman splitting) -- dust grains become aligned and emit
so that the polarization
psuedo-vectors are perpendicular to the magnetic field direction.
As well as the importance of magnetic pressure for star formation
being unclear, details of grain alignment also need more careful comparisons
between observations and modelling.

Studies of magnetic fields in the ISM regions are complicated by the fact
that few observational means exist to characterize their strength and
morphology. However, some of these techniques are well-suited to the submm
window. For years the main tool for finding
the orientation of magnetic fields in the plane of the sky has been to use
using the dispersion of polarization measurements (the
Chandrasekhar-Fermi method), which is only reliable
using data with high enough spatial resolution.
(another area where Canadians are leading [16]).
Polarization of molecular lines such as CO can also yield important
information when dust
emission is weaker.  And another technique compares coexistent
neutral and ion molecular species to probe
magnetic fields in turbulent clouds.

These techniques will help answer several outstanding questions --
How are magnetic fields in spiral arms connected with those in molecular
clouds?  Does magnetism dominate the turbulent energy budget?
What processes align the dust grains?  Is there a difference between large
and small grains?  What is the anomalous emission seen as a foreground in
CMB experiments?

SCUBA-2 POL is a Canadian led enhancement for SCUBA-2, and 
the BLAST experiment has been redesigned as `BLAST-Pol', with polarization
sensitivity.  These facilities will help understand the
relationship between magnetic fields in the ISM and in dense cores.

\begin{figure}
  \begin{center}
    \leavevmode
      \epsfxsize=8cm\epsfbox{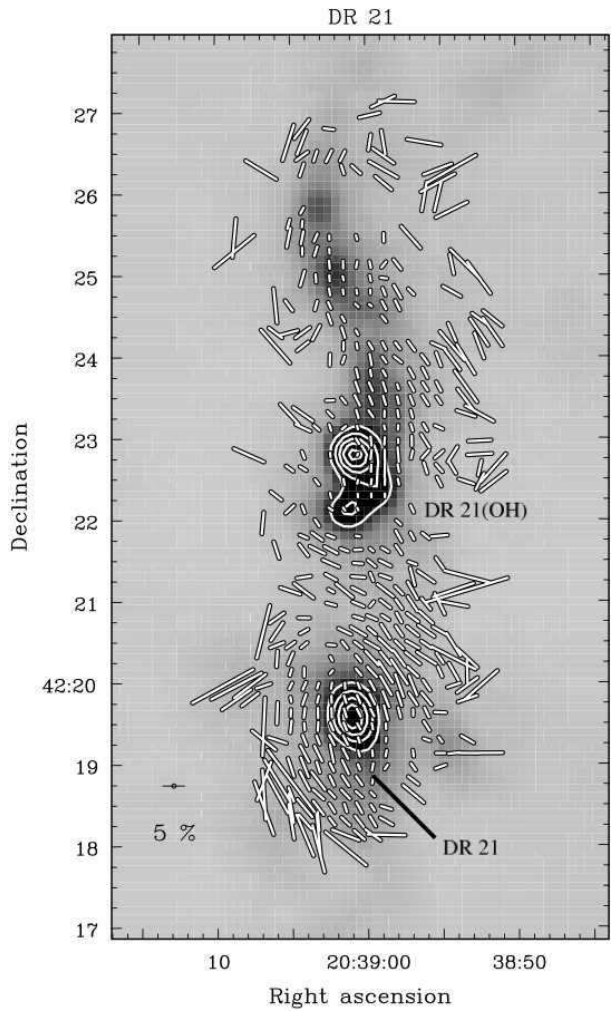}
       \caption{SCUBA $850\,\mu$m image of the high-mass
       star-forming filament DR21 (grey-scale, with contours starting
       at $5\,{\rm Jy}\,{\rm beam}^{-1}$).  Polarization pseudo-vectors
       are overlaid (for pixels with sufficient signal-to-noise), showing
       coherent magnetic field patterns [17].}
     \label{fig:DR21}
  \end{center}
\end{figure}

\begin{figure}
  \begin{center}
    \leavevmode
      \epsfxsize=8cm\epsfbox{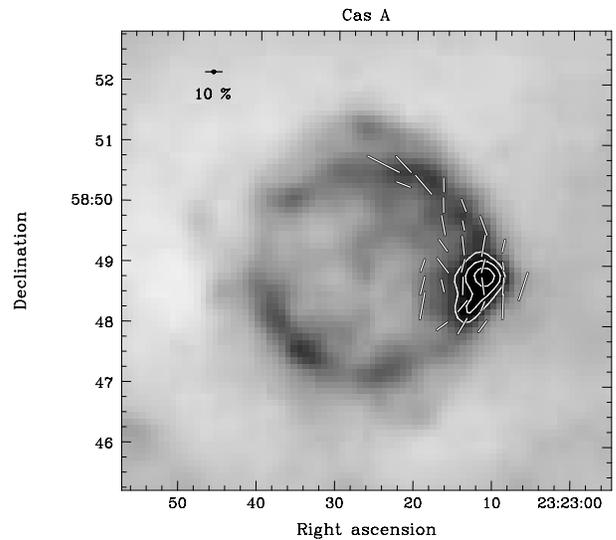}
       \caption{SCUBA $850\,\mu$m imaging and polarimetry of the
       Cas~A supernova remnant [17], showing an ordered pattern in the shell.}
     \label{fig:CasA_POL}
  \end{center}
\end{figure}

\subsection{The Galactic Centre}
The Central Molecular Zone is a region of about $200\,$pc radius around the
centre of the Milky Way Galaxy, containing about $4\times10^7{\rm M}_\odot$
of warm molecular gas and dust.  This is about 10\% of the total ISM in
the Galaxy, and can be
used as a local template for studying gas-rich AGN.
What is the dominant heating process ($T$ is higher than in the Plane)?
What are consequences of the strong magnetic field?
What is going on right down at the scale of the Schwarzschild Radius?

\begin{figure}
  \begin{center}
    \leavevmode
      \epsfxsize=9cm\epsfbox{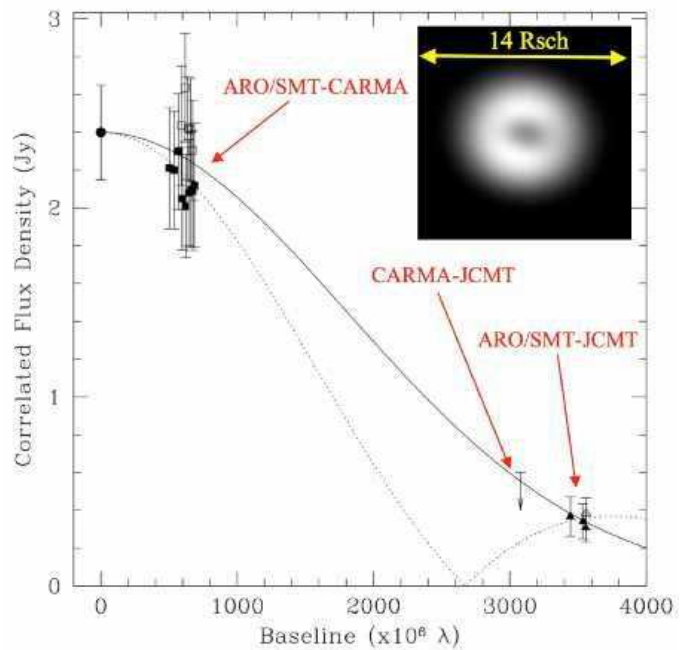}
       \caption{Measured values for the correlated flux density of Sgr~A$^\ast$
       as a function of baseline compared with 2 models.  The solid line is
       for a simple Gaussian of FWHM of $0.037^{\prime\prime}$ (corresponding
       to $3.7\,R_{\rm Sch}$), while the
       dashed line is for the image shown in the inset, which has a
       silhouette in its accretion disk, caused by relativistic effects [19].}
     \label{fig:horizon}
  \end{center}
\end{figure}

In the submm we can probe right into the obscured core of the Galaxy.
Variability and polarimetry studies have already shown promise as tools
for probing the complex dynamics in the Galactic Centre.
Recent VLBI observations show that it is possible
to see down to the event horizon [18], a prospect which will be improved with
the use of ALMA.

\subsection{Planetary Systems and Disks}

The submm also allows us to probe cold disk systems around stars.  Collisions
around some stars have generated debris disks which can be relatively bright
at these wavelengths, enabling us to study analogues of the Solar System's
Kuiper Belt.  Similar studies can be made of genuinely protoplanetary disks,
as well as systems in transition.  SCUBA allowed us to probe the structure of
a few nearby disks (Fig.~8),
while surveys with SCUBA-2 and {\it Herschel\/} will find
new targets for detailed follow-up.  Again this is an area where Canadians
are leading the effort (see the separate White Paper [19]).

ALMA gives us the spatial resolution to probe disk structure in some detail,
including radial variation of the spectral energy distribution.  However,
exploring the chemistry will be more challenging.
An ambitious goal is to resolve water vapour in
planet-forming disks -- this will require the spatial resolution of a far-IR
interferometer.

\begin{figure}
  \begin{center}
    \leavevmode
      \epsfxsize=8cm\epsfbox{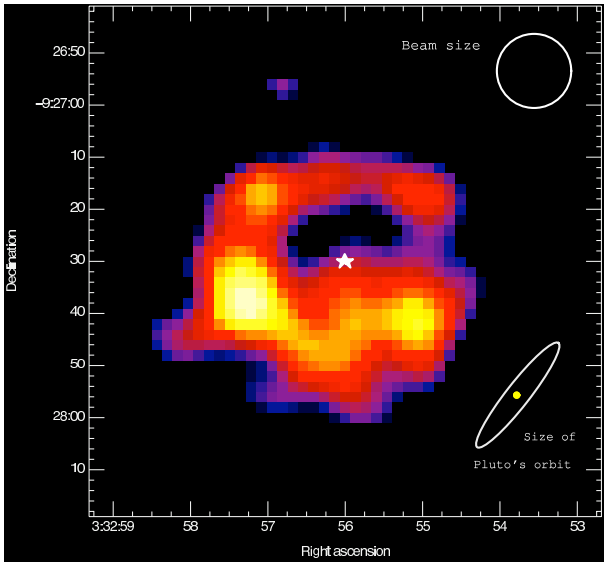}
\vspace{0.3truecm}
       \caption{SCUBA $850\,\mu$m image of dust around the star
       $\varepsilon$ Eridani (marked in the centre).  The extended flux
       resembles a ring of circumstellar dust, perhaps corresponding to the
       Solar System's Kuiper Belt and also having intriguing asymmetries [20].}
     \label{fig:eps_eridani}
  \end{center}
\end{figure}

\subsection{The Solar System}

Submm observations have been used to investigate a diverse range of topics
in our Solar System, including: the kinematic and temperature
structure in terrestrial planetary atmospheres; chemistry in Jovian planets;
albedos of asteroids; and dust and molecular physics in comets.  These
studies are only possible at submm wavelengths.  Future work will involve
using the new facilities, e.g.~ALMA's more meaningful angular resolution, or
increased sensitivity enabling Kuiper Belt objects to be studied (Fig.~9).

\begin{figure}
  \begin{center}
    \leavevmode
      \epsfxsize=8cm\epsfbox{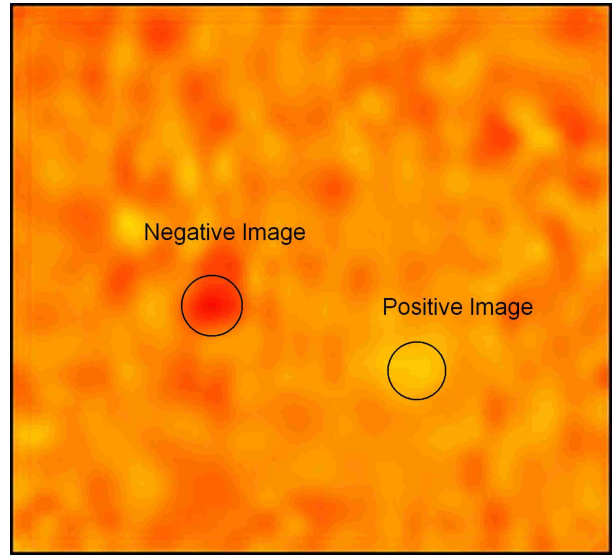}
\vspace{0.3truecm}
       \caption{{\it Herschel}-SPIRE image of the dwarf planet MakeMake, using
       a differencing technique on observations taken tens of hours apart.
       This shows the potential for observing faint Solar System objects, even
       below the background confusion limit at these wavelengths (image
       from ESA and the `TNOs are cool' {\it Herschel\/} Key Project).}
     \label{fig:makemake}
  \end{center}
\end{figure}

\section{Extragalactic Science Goals}

For starbursting galaxies the dust continuum contains the bulk of the
luminosity, and photometry can straightforwardly determine dust temperature
and emissivity.  Lines provide around 1\% of the luminosity, and can be used
to determine physical conditions, as well as to obtain redshifts.
It is important to probe both the gas and the dust, in order to relate the
fuel for future star formation to the emission from current stars.

The Far-IR/submm bands include 10 fine-structure lines from abundant species
-- C{\sc i}, C{\sc ii}, N{\sc ii}, N{\sc iii}, O{\sc i} and O{\sc iii},
plus mid-$J$ rotational transitions of CO (Fig.~10).
Extragalactic astrochemistry is in its infancy, but it is important to
extend Milky Way studies to understand physical processes in nearby galaxies,
and apply this knowledge to more distant Ultra-Luminous Infrared Galaxies
(ULIRGs) and Sub-Millimetre Galaxies (SMGs).
Single dish telescopes with wide-band spectrometers (perhaps CCAT in future)
are useful for studying whole galaxies, while ALMA can resolve molecular cloud
scales.  Space-based instruments, such as SOFIA, {\it Herschel} and SPICA
extend the wavelength (and redshift) range.

\subsection{Nearby Galaxies}

The ability to resolve structures within nearby galaxies enables detailed
studies to be carried out over a 
wider range of star-forming environments than are accessible
in the Milky Way.  Super-star clusters in galaxy mergers, for example,
can help to understand high-$z$ galaxy formation.  Specific molecular
transitions probe quiescent gas, grain chemistry in photo-warmed gas, shocks,
dense gas clumps and higher temperature gas.  Questions that can
be addressed include:
How are star formation estimates from submm emission related to estimates from
other wavebands (see [21])?
What is the relationship between the global chemical
structure of galaxies and their dynamical history?
How do starbursts feed back on molecular clouds?
How good is HCN as a star formation tracer?
Can chemistry distinguish AGN from starburst luminosity?
How does dust grain size and composition vary with environment and time?

\begin{figure}
  \begin{center}
    \leavevmode
      \epsfxsize=8cm\epsfbox{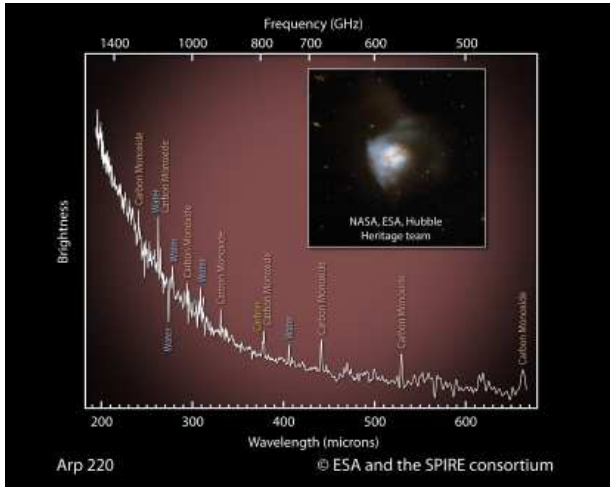}
\vspace{0.3truecm}
       \caption{An early result from the {\it Herschel}-SPIRE Fourier
       Transform Spectrometer.  This is a sub-mm spectrum of the relatively
       nearby starbursting galaxy Arp220, showing many individual lines --
       part of the `Nearby Galaxies' Key Programme, in which
       Canadians are playing a leading role.}
     \label{fig:Arp220_spectrum}
  \end{center}
\end{figure}

\subsection{The Peak of Star Formation}

Submm observations can be used to determine global star formation history,
and distinguish among models of galaxy formation.
Star formation per unit comoving volume peaks when the Universe was
perhaps 30\% of its present age ($z\,{=}\,1$--2).  Galaxies with the highest
star formation are also the dustiest, so often impossible to directly
view in the optical.  Long past their era of activity, SMGs or ULIRGs peak
around $z\,{\simeq}\,2.5$ (Fig.~11), and although rare they are extremely
luminous, so they are a significant contributor to the global star-formation
rate at high $z$.

\begin{figure}
  \begin{center}
    \leavevmode
      \epsfxsize=8cm\epsfbox{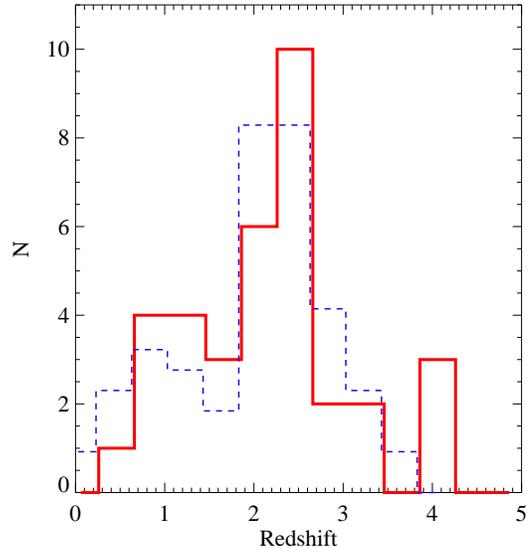}
       \caption{Redshift distribution of Submillimetre Galaxies
       (SMGs) [22].
       The solid histogram is from the nearly complete GOODS-North sample,
       while the dashed curve is the larger sample of radio selected SMGs
       [23].  The average redshift appears to be $z\,{\simeq}\,2.5$, with
       little evidence for a significant tail beyond $z\,{=}\,4$.}
     \label{fig:red_dist}
  \end{center}
\end{figure}

ULIRGs are clearly related to mergers, as are AGN, but the exact relationship
is currently unknown -- how much of the rest-frame far-IR emission has a
gravitational rather than nuclear origin?  And do ULIRGs turn into quasars?

Theoretical understanding of dark matter is under control, but the
baryonic mass assembly history of galaxies is much more complicated.
Current submm telescopes allow us to probe the most extreme star-formers,
and future facilities like ALMA (Fig.~12) will enable us to observe normal
Milky Way-like galaxies at high $z$, not just the monsters.

{\it Herschel\/} has limited ability to carry out extragalactic spectroscopy
-- even the brightest far-IR lines redshifted into the submm are extremely faint
for all but the most luminous galaxies at $z\,{>}\,1$.  So progress out to the
highest redshifts will require the additional sensitivity provided through
the use of a cooled telescope -- SPICA, the 3.5--m cryogenically cooled
Japanese-led telescope, is planned for launch in 2017.

\subsection{The First Galaxies}

The negative K-correction in the submm (Fig.~13)
ensures that as soon as star-forming galaxies
contain dust, they will be relatively easy to detect, even at
$z\,{\sim}\,10$.  Additionally, redshifted far-IR lines such as from C{\sc ii}
provide sharper diagnostics.  Can we use these observations to help determine
when the first stars and accretion-powered black holes formed?  And at what
epoch was the star formation {\it not\/} dust-enshrouded?
Can we trace the end of the Dark Ages through cooling lines as the Universe
transitions from a dustless to dusty medium?  The sensitivity
afforded by ALMA and eventually SPICA will help perform spectroscopic studies
of the earliest objects.

\begin{figure}
  \begin{center}
    \leavevmode
      \epsfxsize=8cm\epsfbox{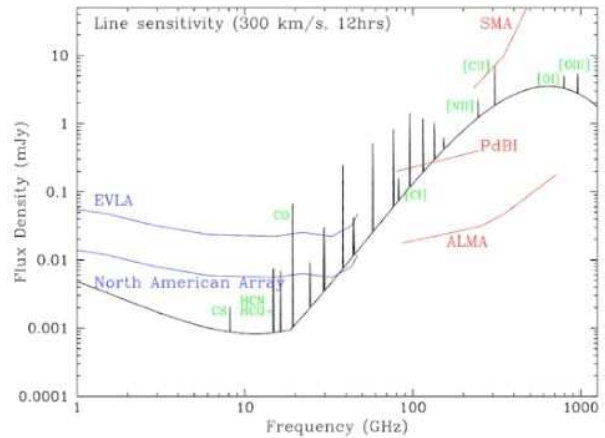}
       \caption{Model spectrum of the starburst galaxy Arp220, shifted
       to $z\,{=}\,5$ and with several molecular and atomic lines labelled
       [26].  The red lines show the $1\sigma$ sensitivities
       for a 12 hour observation of a $300\,{\rm km}\,{\rm s}^{-1}$ linewidth
       for the 3 submm interferometers: Submillimeter Array; IRAM Plateau
       de Bure; and Atacama Large Millimeter/submillimeter Array.}
     \label{fig:Arp220}
  \end{center}
\end{figure}

\begin{figure}
  \begin{center}
    \leavevmode
      \epsfxsize=8cm\epsfbox{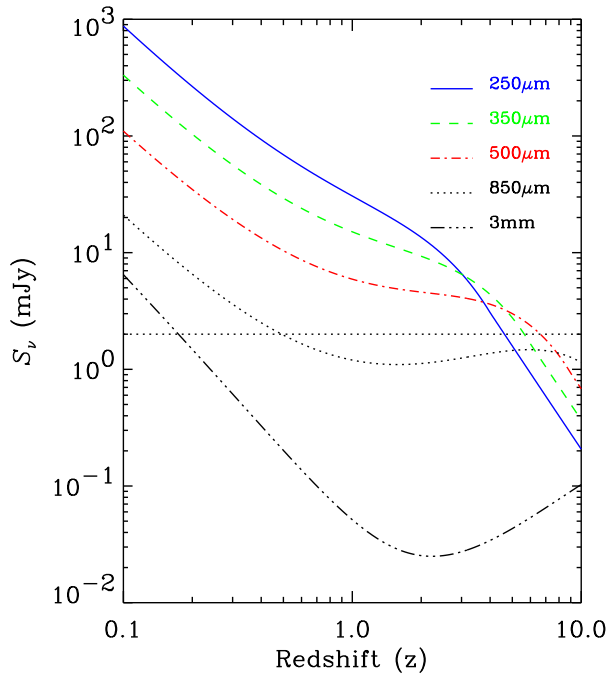}
       \caption{Galaxies of a given luminosity are almost equally
       detectable over a very wide redshift range at submm wavelengths.  This
       is because of the so-called `negative K-correction' pushing the
       observed band {\it up\/} the Rayleigh-Jeans part
       of the spectral energy distribution as the galaxy redshifts,
       compensating for the redshift dimming.  This figure shows the spectrum
       of Arp220 redshifted in several different wavebands [24].
       The horizontal line is
       at $2\,$mJy, approximately the limiting $850\,\mu$m flux density for
       SCUBA.}
     \label{fig:k_corr}
  \end{center}
\end{figure}

\subsection{The Extragalactic Background}

The Cosmic Infrared Background is (probably) entirely composed of
individual galaxies.  A major current endeavour is to resolve the background
and determine the luminosity function and evolution of the galaxies which
comprise it.  This becomes more difficult at longer wavelengths, where there
is a greater contribution from high redshift sources, and where source
confusion is increasingly severe (Fig.~14).

Precise measurements of the background and comparisons with sources
can determine if there are missing components.  Correlations in the background
yield additional information about galaxy evolution within clustered dark
matter haloes.  This will be a major topic for wide surveys, with
{\it Herschel}-SPIRE and SCUBA-2, for example.  One unique aspect of SCUBA-2
is that the resolution available at $450\,\mu$m allows for most of the
background to be resolved directly into sources down to `normal'
$L_\ast$ galaxies (see Fig.~15).

\begin{figure}
  \begin{center}
    \leavevmode
      \epsfxsize=8cm\epsfbox{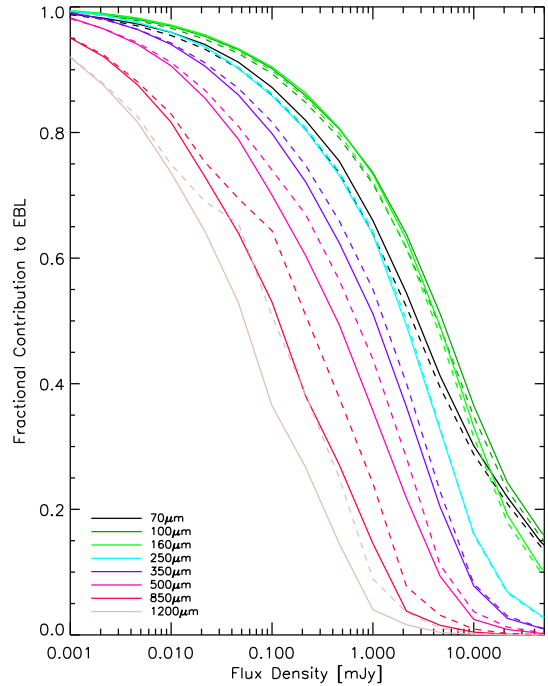}
       \caption{Model estimates for the fractional contribution to
       the extragalactic background from sources above a given flux density at
       different wavelengths (from [25]).  The solid and dashed
       curves represent two different luminosity functions which are consistent
       with most of the current observational constraints.  This shows the need
       for surveys using large dishes (small beamsizes), such as with JCMT or
       CCAT, since the fraction resolved at the confusion limit of SPIRE
       (${>}\,10\,$mJy) is small.
       ALMA has the resolution to detect almost everything which
       contributes, but with small samples because of the small field of view.}
     \label{fig:resolved}
  \end{center}
\end{figure}

\begin{figure}
  \begin{center}
    \leavevmode
      \epsfxsize=9cm\epsfbox{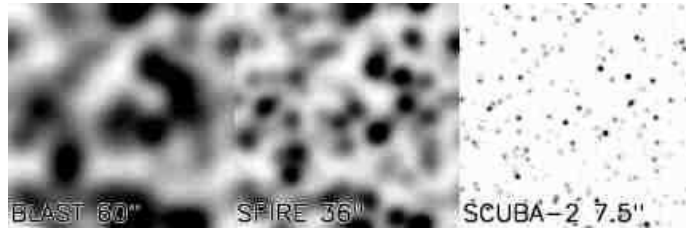}
       \caption{Comparison of the effects of the
       beamsizes of BLAST, SPIRE and SCUBA-2 at $450\,\mu$m in a deep
       extragalactic simulation.  While great progress can be made through
       statistical studies of confused maps, SCUBA-2 will detect objects
       individually.}
     \label{fig:rescompare}
  \end{center}
\end{figure}

\subsection{AGN and black holes}
As well as probing the first galaxies, we would also like to understand
when the first super-massive black holes formed and determine the conditions
which drive their evolution (see [27]).  The submm band provides a
means of studying the gaseous fuel through atomic fine structure lines and
molecular transitions.   High resolution spectroscopic imaging can
probe galaxy dynamics and star-formation processes on sub-kpc scales, seeing
through the dust which hampers most other wavebands.

\begin{figure}
  \begin{center}
    \leavevmode
      \epsfxsize=8cm\epsfbox{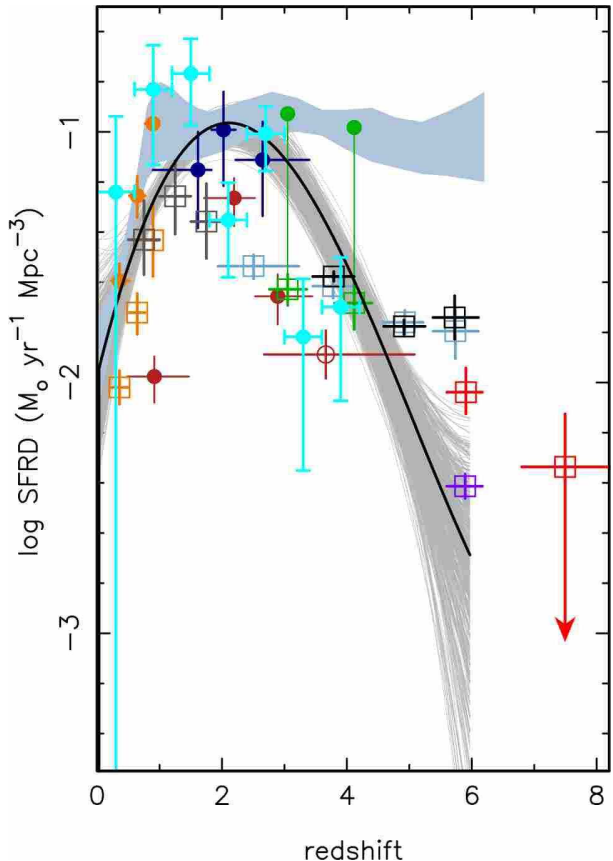}
       \caption{Star-formation rate density determined from a
       sample of SMGs (cyan points) compared with estimates from other
       methods (see [30]).  The solid line and shading
       represent the space density of radio QSOs (arbitrarily normalized),
       showing the similarity between the history of star formation and black
       hole accretion.}
     \label{fig:SMG_evolution}
  \end{center}
\end{figure}

The connection between star formation and AGN activity is still a great
mystery (Fig.~16).  Mergers provide a natural link, but details
depend on physical processes which are hard to model.  Moreover, the feedback
of the AGN onto star formation is also poorly understood.  Spectroscopically
with {\it Herschel}, ALMA and SPICA are again just what is needed.

\subsection{Galaxy Clusters}

How much total power is emitted by galaxies in different environments?
What types of galaxies dominate global star formation, and how do stars
form within them?
When and how did galaxies form and their metals accumulate?  Do high
redshift star-forming galaxies avoid overdensities like we see locally or is
there a reversal in this relation (see also [29])?
These and other questions involving the relationship between galaxies and
clusters are
only now becoming possible because of the sensitivity of modern submm
facilities.

The $850\,\mu$m atmospheric window is close to the peak of the
Sunyaev-Zel'dovich (SZ) {\it increment}.  Combination with more traditional
measurements of the decrement improve the separation of the
SZ effect from radio and star-forming galaxies within the clusters,
as well as lensed background sources, potentially yielding
estimates of the kinetic SZ effect to study cluster bulk velocities.
Calibrating properties of
galaxy clusters for use as cosmological tools requires comprehensive study,
and here the submm observations will certainly help.  In addition
the high resolution provided by ALMA will enable detailed investigation
of intracluster physics through the thermal SZ
effect from substructure.

It is well-known that rich galaxy clusters can act as lenses for background
SMGs.  The steep number counts mean that we can use
`natures telescopes' to probe down to
intrinsically much fainter objects at high-$z$.  Hence wide surveys may uncover
strong lensing candidates more efficiently than radio and optical surveys
(see [30]).
Once found, follow-up observations of such lenses can be used
to probe dark matter potential wells, including sub-structure, as well as
constraining dark matter physics and the influence of super-massive black
holes.

\begin{figure*}
  \begin{center}
    \leavevmode
      \epsfxsize=15cm\epsfbox{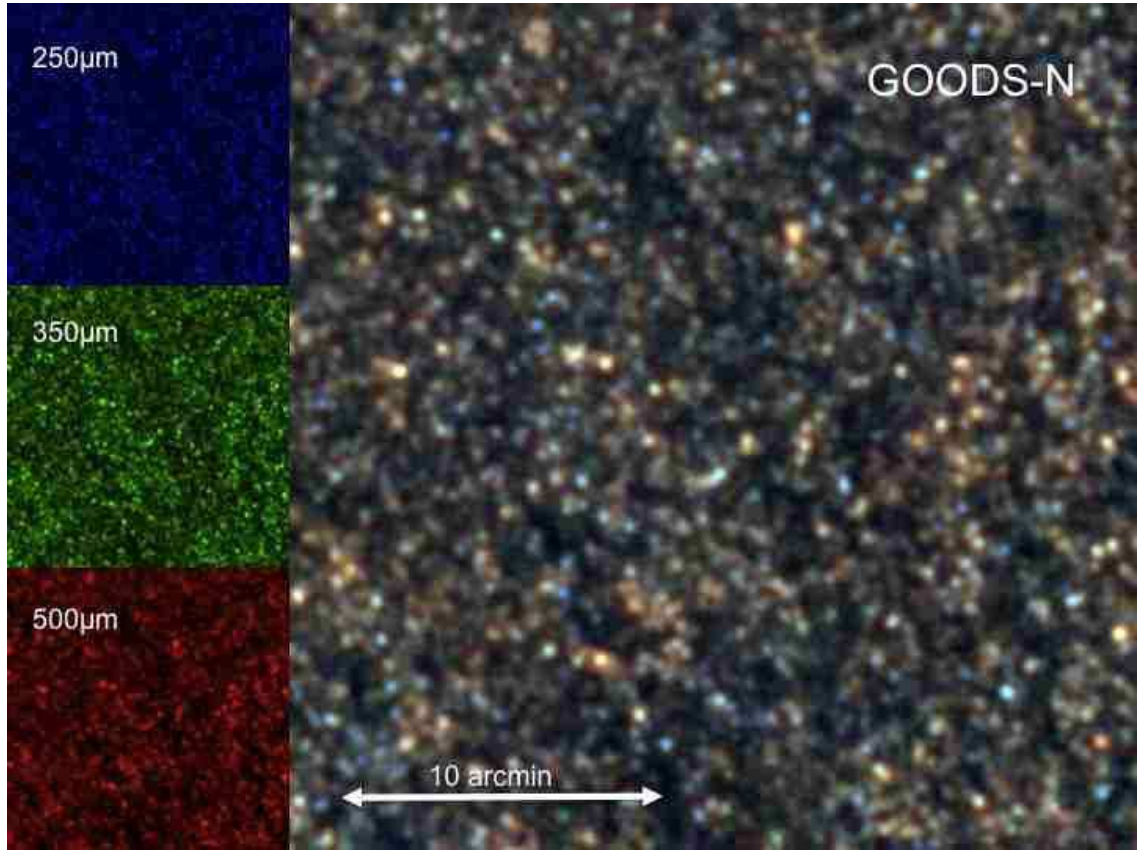}
       \caption{Composite colour image of the Great Observatories
       Origins Deep Survey -- North field.  This was imaged using
       {\it Herschel\/}-SPIRE as part of the HerMES guaranteed time project.
       Cite HerMES web-site [33], this particular figure put together by
       Gaelen Marsden for the HerMES press release.  It shows the quality of
       SPIRE data, and the coming of age of extragalactic submm astronomy
       -- this single image contains thousands of sources.}
     \label{fig:GOODSN}
  \end{center}
\end{figure*}

\section{Recommendations}

A wide range of science questions can be tackled using unique diagnostics
at submm wavelengths.  Most of them
require deep and wide surveys encompassing tens to thousands of square
degrees and tens of thousands to millions of galaxies.  Detailed follow-up of
targets demands high spatial resolution and sensitive spectroscopy.
As a consequence of these drivers it is natural that a combination of
facilities is required.

In addition to instrumental facilities, there are also other factors
which will enable progress.  Map-making at submm wavelengths
(using iterative techniques to approach the maximum likelihood solution)
has become computationally intensive, as has source extraction (using
multi-wavelength Bayesian techniques for example), statistical analysis of
images (e.g.~with so-called $P(D)$ analysis) and comparison with models
(including hydrodynamic galaxy mergers or turbulent ISM simulations).
As specific examples, both BLAST and SCUBA-2 use cluster computing and
have developed archiving plans through the Canadian Astronomy Data
Centre (see [31] and [32]).  A second factor is human resources --
Canadians are behind most of their international partners in the ability
to fund the postdoctoral researchers who can carry the complex analysis to
fruition.

Among Canadian astronomers with submm interests there is a high degree of
consensus in the following specific recommendations:

\begin{enumerate}
\item Complete JCMT Surveys
\item Vigorous involvement in {\it Herschel}
\item Prepare for ALMA
\item Develop plans for involvement in future survey instrument(s)
\item Plan for future space-based mission(s)
\item Support theoretical work
\end{enumerate}

A decade ago we were struggling to make maps which encompassed more than a
few star-forming cores or distant SMGs.  With the coming of BLAST (Fig.~3),
{\it Herschel\/} (see Fig.~17) and SCUBA-2 we have moved well beyond the
`one object at a time' science, and can now survey thousands of sources at
once.  Continuing this survey work and performing targeted follow-up with
ALMA and other facilities means that we have a very productive decade ahead
of us.

\acknowledgements
This White Paper is based on results from a huge number of
papers and discussions with many of our colleagues.
Due to the review nature of this contribution we have had to omit
most original citations, but most can be found by tracing through the other
reviews to which we refer.

\end{document}